\documentclass[a4paper]{article}
\usepackage[super,sort&compress]{natbib}
\usepackage{epsfig,lineno}
\usepackage[affil-it]{authblk}

\bibpunct{}{}{,}{s}{,}{,}
\fussy
\begin{document}

%\linenumbers

\title{Stochastic Climate Theory and Modelling}

\author[1]{Christian L. E. Franzke\thanks{Corresponding Author;
    Meteorological Institute and Centre for Earth System Research and
    Sustainability, University of Hamburg, Grindelberg 7, D-20144 Hamburg,
    Germany; Email: christian.franzke@uni-hamburg.de}}

\author[2]{Terence J. O'Kane}

\author[3]{Judith Berner}

\author[4]{Paul D. Williams}

\author[1,5]{Valerio Lucarini}

\affil[1]{Meteorological Institute and Centre for Earth System
  Research and Sustainability (CEN), University of Hamburg, Hamburg, Germany}
\affil[2]{Centre for Australian Weather and Climate Research, CSIRO
  Marine and Atmospheric Research, Hobart, Australia}
\affil[3]{National Center for Atmospheric Research, Boulder, USA}
\affil[4]{Department of Meteorology, University of Reading, Reading,
  UK}
\affil[5]{Department of Mathematics and Statistics, University of Reading,
  Reading, UK}

\date{\today}

\maketitle

\begin{center}
\noindent{\bf Abstract}
\end{center}
Stochastic methods are a crucial area in contemporary climate research and
are increasingly being used in comprehensive weather and climate prediction
models as well as reduced order climate models. Stochastic methods are used as
subgrid-scale parameterizations as well as for model error representation,
uncertainty quantification, data assimilation and ensemble
prediction. The need to use stochastic approaches in weather and
climate models arises because we still cannot resolve all necessary
processes and scales in comprehensive numerical weather and climate
prediction models. In many practical applications one is mainly
interested in the largest and potentially predictable scales and not
necessarily in the small and fast scales. For instance, reduced order
models can simulate and predict large scale modes. Statistical
mechanics and dynamical systems theory suggest that in
reduced order models the impact of unresolved degrees of freedom can
be represented by suitable combinations of deterministic and
stochastic components and non-Markovian (memory) terms. Stochastic approaches
in numerical weather and climate prediction models also lead to the
reduction of model biases. Hence, there is a clear need for systematic
stochastic approaches in weather and climate modelling. In this review
we present evidence for  stochastic effects in laboratory
experiments. Then we provide an overview of stochastic climate theory
from an applied mathematics perspectives. We also survey the current
use of stochastic methods in comprehensive weather and climate
prediction models and show that stochastic parameterizations have the
potential to remedy many of the current biases in these comprehensive
models.

\clearpage

\section{Introduction}
The last few decades have seen a considerable increase in computing
power which allows the simulation of numerical weather and climate prediction
models with ever higher resolution and the inclusion of ever more physical
processes and climate components (e.g. cryosphere, chemistry). Despite
this increase in computer power many important physical processes
(e.g. tropical convection, gravity wave drag, clouds) are still not
or only partially resolved in these numerical models. Despite the
projected exponential increase in computer power these processes will not be
explicitly resolved in numerical weather and climate models in the
foreseeable future\cite{Williams:2005b,Palmer:2008}. For instance, Dawson et al.
\cite{Dawson:2012} have demonstrated using the ECMWF integrated forecast system that
extremely high resolutions (T1279, which corresponds to a grid spacing
of about 16km) are required to accurately simulate the observed Northern
hemispheric circulation regime structure. This resolution, typical
for limited area weather and climate models used for short term prediction,
remains unfeasible for the current generation of high resolution global
climate models due to computational and data storage requirements. Hence,
these missing processes need to be parameterized, i.e. they need to be
represented in terms of resolved processes and scales
\citep{Stensrud:2007}. This representation is important because small-scale
(unresolved) features can impact the larger (resolved) scales
\citep{Lorenz:1969,Tribbia:2004} and lead to error growth, uncertainty and 
biases.

At present, these parameterizations are typically deterministic, relating
the resolved state of the model to a unique tendency representing the integrated effect
of the
unresolved processes. These ``bulk parameterizations'' are based on the notion
that the properties of the unresolved subgrid-scales are determined by
the large-scales. However, studies have shown that resolved states
are associated with many possible unresolved states
\citep{ Crommelin:2008, ShPa07,Wilks:2005}. This calls for stochastic methods
for numerical weather and climate prediction which 
potentially allow a proper representation of the uncertainties, a reduction of
systematic biases and improved representation of long-term climate
variability. Furthermore, while current deterministic parameterization schemes are
inconsistent with the observed power-law scaling of the energy
spectrum \cite{Sh05,Be09} new statistical dynamical approaches
that are underpinned by exact stochastic model representations have emerged
that overcome this limitation. The observed power spectrum structure is
caused by cascade processes. Recent theoretical studies suggest that
these cascade processes can be best represented by a stochastic
non-Markovian Ansatz. Non-Markovian terms are necessary to model memory
effects due to model reduction \cite{Chorin:2013}. It means that in order to make skillful
predictions the model has to take into account also past states and not only
the current state (as for a Markov process).

We first review observational evidence of stochasticity in laboratory
geophysical fluid experiments (section 2), then discuss stochastic
climate theory in fast-slow systems (system 3). In section 4 we
present statistical physics approaches and in section 5 we review the
current state of stochastic-dynamic weather and climate modelling. We close
with an outlook and challenges for the future of weather and climate
modelling (section 6).

\section{Laboratory Evidence of Stochasticity}
Research on the climate system is somewhat hindered by the obvious
difficulties of performing reproducible experiments on
the atmosphere and ocean in different parameter regimes. For example, an optical
physicist studying the nonlinear response of isolated atoms to intense
electromagnetic waves can easily change the incident wavelength
\cite{Okane:1999}.  In contrast, climate scientists cannot (and
arguably should not!) change the rotation rate of the planet or the
intensity of the incoming solar radiation. To some extent, numerical
simulations come to the rescue, by allowing us to perform virtual
experiments. However, the grid spacing in climate models is orders of
magnitude larger than the smallest energized scales in the atmosphere
and ocean, introducing biases.

Fortunately, there is another option available to us. It is possible
to exploit dynamical similarity \cite{Douglas:2000} to study analogues of
planetary fluid flow in bespoke laboratory experiments. The
traditional set-up is the classic rotating annulus, which has been
used for decades to study baroclinic instability and other large-scale
phenomena \cite{Hide:1977}. Recent observations of small-scale
inertia-gravity waves embedded within a large-scale baroclinic wave
\cite{Lovegrove:2000,Williams:2005,Williams:2008} have allowed the
scale interactions between these two modes to be studied in a
laboratory fluid for the first time. The experimental apparatus
consists of a two-layer isothermal annulus forced by a differentially
rotating lid, which drives a shear across the internal interface and
represents the mid-latitude tropospheric wind shear.

\begin{figure}
\begin{center}
\noindent\includegraphics[width=1.0\textwidth]{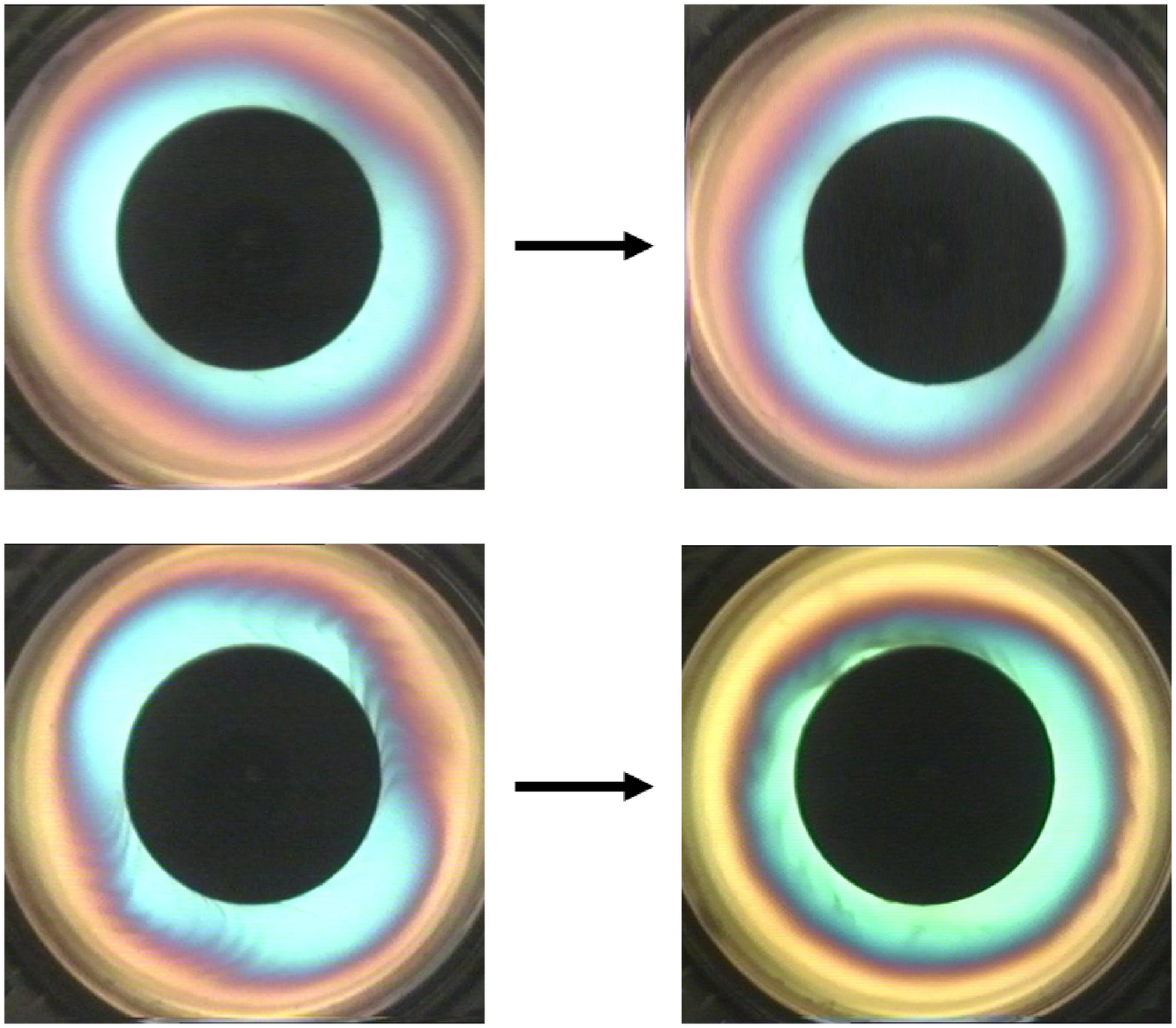}
\end{center}
\caption{\label{fig:transitions}
Regime transitions in a rotating two-layer annulus laboratory
experiment, viewed from above.  Different colours correspond to
different internal interface heights, through the use of a
sophisticated visualisation technique \cite[]{Williams:2004b}.  In the
upper row, small-scale inertia--gravity waves are absent, and
large-scale regime transitions do not occur.  In the lower row, small-scale
inertia--gravity waves are present locally in the troughs of the
large-scale wave, and a large-scale regime transition does occur.  From the laboratory
experiments of Williams et al.
\cite{Williams:2003,Williams:2004,Williams:2005,Williams:2008}.}
\end{figure}

The large-scale baroclinic wave in these laboratory experiments
exhibits regime behavior, equilibrating at finite amplitude with a
zonal wavenumber of typically 1, 2, or 3.  These simple wave modes are
regarded as prototypes of the more complicated regime behavior in the
atmosphere, such as mid-latitude blocking \cite[]{Weeks:1997,Tian:2001}.  A
notable finding from repeated experiments using this apparatus is
that small-scale inertia--gravity waves can induce large-scale
regime transitions, despite the separation of wavelengths by an order
of magnitude \cite[]{Williams:2003}.  An example of this process is
illustrated in Figure~\ref{fig:transitions}.  A wavenumber 2 mode {\em
  without} co-existing inertia-gravity waves (upper row) remains
a wavenumber 2 mode indefinitely, drifting around the annulus with the
zonal-mean flow.  In contrast, with the same parameter values, a
wavenumber 2 mode {\em with} co-existing inertia--gravity waves (lower
row) is found to have a finite probability of transitioning to a
wavenumber 1 mode.  The amplitude of the inertia--gravity waves is
controlled here without directly affecting the large-scale mode, by
slightly varying the interfacial surface tension between the two
immiscible fluid layers.

The laboratory transitions discussed above are reminiscent of
noise-induced transitions between different equilibrium states in a
meta-stable dynamical system \cite{Swart:1987}. To test this
interpretation, a quasi-geostrophic numerical model that captures the
meta-stability of the large-scale flow in the rotating annulus
\cite[]{Williams:2009} was run with and without weak stochastic forcing added
to the potential vorticity evolution equation for each fluid layer.
The stochastic forcing was an approximate representation of the
inertia--gravity waves, which are inherently ageostrophic and are
therefore forbidden from the quasi-geostrophic model. Consistent with
the laboratory experiments, only when the noise term was activated did
the numerical simulations exhibit large-scale wave transitions in the
equilibrated flow \cite[]{Williams:2004}. In further numerical experiments,
the noise was found to be able to influence wavenumber selection
during the developing baroclinic instability.

In summary, the above laboratory experiments constitute the first
evidence in a real fluid that small-scale waves may trigger
large-scale regime transitions. In a numerical model in which the
small-scale waves were absent, the transitions were captured through
the addition of stochastic noise. Note that the small-scale waves
satisfy the dispersion relation for inertia--gravity waves and are
therefore coherent in space and time, and yet apparently they are
'sensed' by the large-scale flow as if they were random fluctuations.
These results have led to a possible interpretation of sudden
stratospheric warmings as noise-induced transitions
\cite[]{Birner:2008}. Furthermore, these laboratory results help to motivate
the development of stochastic parameterizations in climate models and
a more general development of stochastic climate theory.

\section{Stochastic Climate Theory}
Climate is a multi-scale system in which
different physical processes act on different temporal and spatial
scales \cite{Klein:2010}. For instance, on the micro-scale are turbulent eddies with
time scales of seconds to minutes, on the meso-scale is convection
with time scales of hours to days, on the synoptic scale are
mid-latitude weather systems and blocking with time scales from days
to weeks, on the large-scale are Rossby waves and teleconnection
patterns with time scales of weeks to seasons. And there is the
coupled atmosphere-ocean system with time scales of seasons to
decades. The crucial point here is that all these processes acting on
widely different temporal and spatial scales, interact with each
other due to the inherent nonlinearity of the climate system. We have
shown an illustrative laboratory example for this in the previous
section.

For many practical applications we are only interested in the
processes on a particular scale and not in the detailed evolution of
the processes at smaller scales. Often the scales of interest are
linked to inherently predictable processes, while the smaller scales
processes are unpredictable. For instance, in the above laboratory
experiment we are interested in the regime behavior and not in the
detailed evolution of the inertia--gravity waves. In numerical
simulations the fastest scales, which are typically also the smallest
scales, use up the bulk of computing time, slowing down the
computation of the processes of actual interest. In numerical weather
and climate prediction many of the small scale processes are currently
not explicitly resolved and won't be in the foreseeable future. This
neglect of these processes can lead to biases in the
simulations. Because of that the unresolved processes need to be
parameterized as demonstrated in the previous section.

Stochastic climate theory is based on the concept of scale separation
in space or time. \citet{Hasselmann:1976} was the first to propose to split
the state vector $\vec{z}$ into slow climate components $\vec{x}$ and fast
weather fluctuations $\vec{y}$ and then to derive an effective equation for the
slow climate variables only. In this equation the effect of the now
unresolved variables is partially represented as a noise term. The physical intuition
behind this idea is, for example, that the aggregated effect of 'fast' weather
fluctuations drives fluctuations in the 'slower' ocean circulation. To first
order such a model can explain the 'red' spectrum of oceanic variables
\citep{Frankignoul:1977,Imkeller:2001}. It has to be noted that there
is no scale separation in the climate system. This lack of time-scale
separation introduces non-Markovian (memory) effects and complicates the
derivation of systematic parameterizations.

Rigorous mathematical derivations for this approach have been provided by
\citet{Khasminsky:1963,Kurtz:1973,Papanicolaou:1976,Pavliotis:2008,Melbourne:2011,Gottwald:2013b}. For
accessible reviews see \citet{Givon:2004} and the text book by
\citet{Pavliotis:2008}. This approach has been applied to climate models by
Majda and coworkers 
\citep{Majda:1999,Majda:2001,Majda:2002,Majda:2003,Majda:2005,Majda:2008,Majda:2009,Franzke:2005,Franzke:2006,Dolaptchiev:2013}. Climate
models have the following general functional form
\begin{equation}
d\vec{z} = \left( \tilde{F} + \tilde{L} \vec{z} + \tilde{B}(\vec{z},\vec{z}) \right) dt
\label{EQ:1}
\end{equation}
where $\tilde{F}$ denotes an external forcing, $\tilde{L}$ a linear
operator and $\tilde{B}$ a quadratic nonlinear
operator. Eq. (\ref{EQ:1}) constitutes the form of the dynamical cores
of weather and climate prediction models.

Now splitting the state vector $\vec{z}$ into slow $\vec{x}$ and fast
$\vec{y}$ components (which amounts to assuming a time scale separation) and
assuming that the nonlinear self-interaction 
of the fast modes $\tilde{B}(\vec{y},\vec{y})$ can be represented by a
stochastic process
\citep{Majda:1999,Majda:2001,Franzke:2005,Franzke:2006} leads to a
stochastic differential equation. The stochastic mode reduction
approach \citep{Majda:1999,Majda:2001,Franzke:2005,Franzke:2006} then
predicts the functional form of reduced climate models for the slow
variable $\vec{x}$ alone:
\begin{equation}
d\vec{x} = \left( F + L \vec{x} + B(\vec{x},\vec{x}) +
  M(\vec{x},\vec{x},\vec{x}) \right) dt + \sigma_A d\vec{W}_A +
\sigma_A(\vec{x}) d\vec{W}_M
\end{equation}
Structurally new terms are a deterministic cubic term which acts predominantly
as nonlinear damping and both additive and multiplicative (state-dependent)
noise terms. The fundamentals of stochastic processes and calculus are
explained in Box 1. The multiplicative noise and the cubic term stem
from the nonlinear interaction between the resolved and unresolved
modes \cite{Franzke:2005}. 

The above systematic procedure allows also a physical interpretation of the
new deterministic and stochastic terms \cite{Franzke:2005}. The additive noise
stems both from the nonlinear interaction amongst the unresolved modes and the
linear interaction between resolved and unresolved modes \cite{Franzke:2005}.

\noindent\fbox{
\parbox{\textwidth}{
{\bf BOX 1}\\
STOCHASTIC PROCESSES\\
In contrast to deterministic processes stochastic processes have a
random component. See the books by \citet{Lemons:2002} and
\citet{Gardiner:1985} for intuitive introductions to
stochastic processes. Typically stochastic processes are driven by white
noise. White noise is a serially uncorrelated time series with zero mean and
finite variance \cite{Gardiner:1985}.

A stochastic differential equation (SDE) is a combination of a deterministic
differential equation and a stochastic process. In contrast to regular
calculus, stochastic calculus is not unique; i.e. different discretizations of
its integral representation lead to different results even for the same noise
realization. The two most important calculi are Ito and Stratonovich. See
details in Gardiner \cite{Gardiner:1985}. The physical difference is that Ito
calculus has uncorrelated noise forcing while Stratonovich allows for finite
correlations between noise increments. Hence, physical systems have to be
typically approximated by Stratonovich SDEs. On the other hand, it is
mathematically straightforward to switch between the two calculi. So one only
needs to make a decision at the beginning which calculus is more appropriate
for modeling the system under consideration and can then switch to the
mathematically more convenient form.

SDEs describe systems in a path wise fashion. The Fokker-Planck equation (FPE)
describes how the probability distribution evolves over time
\cite{Gardiner:1985}. Thus, SDEs and the FPE offer two different ways at
looking at the same system. The parameters of SDEs and their corresponding FPE
are linked; thus, one can use the FPE to estimate the parameters of the
corresponding SDE \cite{Berner:2005,Siegert:1998}.
}
}

Multiplicative (or state-dependent) noise is important for deviations from
Gaussianity and thus extremes. The intuition behind multiplicative noise is as
follows: On a windless day the fluctuations are very small, whereas on a windy
day not only is the mean wind strong but also the fluctuations around this
mean are large; thus, the magnitude of the fluctuations dependent on the state
of the system.

The first practical attempts at stochastic climate modelling were made
using Linear Inverse Models (LIM)

\citep{Penland:1994,Penland:1995,Whitaker:1998,Winkler:2001} and
dynamically based linear models with additive white
noise forcing
\citep{Farrell:1993,Farrell:1995,DelSole:1999,DelSole:2004,Zhang:1999}. These
approaches linearise the dynamics and then add white noise and damping
\citep{Whitaker:1998} in order to make the models numerically stable
(i.e. the resulting linear operator should only have negative
eigenvalues to ensure stability and reliasability of the
solutions). While these models have encouraging predictive skill,
especially for ENSO, they can only produce Gaussian statistics and,
thus, are less useful for predictions of high impact weather.

Recently there are encouraging attempts in fitting nonlinear stochastic
models to data. These include multi-level regression
\citep{Kravtsov:2005,Kondrashov:2006}, fitting the parameters via the
Fokker-Planck equation \citep{Siegert:1998,Berner:2005}, stochastic averaging
\cite{Culina:2011,Monahan:2011}, optimal prediction
\cite{Chorin:2002,Stinis:2006} or Markov Chains
\citep{Crommelin:2006}. Most of the previous approaches fitted the parameters of the
stochastic models without taking account of physical constraints,
e.g. global stability. Many studies linearized the dynamics and then
added additional damping to obtain numerically stable models
\citep{Whitaker:1998,Achatz:1999,Achatz:2003,Zhang:1999}. \citet{Majda:2009}
developed the nonlinear normal form of stochastic climate models and also
physical constraints for parameter estimation. Recent studies use these physical
constraints to successfully derive physically consistent stochastic climate
models \citep{Majda:2013,Harlim:2013,Peavoy:2014}.

Most of the above approaches are based on an implicit assumption of
time scale separation. However, the climate system has a spectrum with
no clear gaps which would provide the basis of scale separation and
the derivation of reduced order models. Such a lack of time scale
separation introduces memory effects into the truncated
description. Memory effects mean that the equations become non-Markovian and
that also past states need to be used in order to predict the next state. This can
be explained by considering the interaction 
between a large-scale Rossby wave with a smaller scale synoptic
wave. At some location the Rossby wave will favor the development of
the synoptic wave. Initially this synoptic wave grows over some days
without affecting the Rossby wave. Once the synoptic wave has reached
a sufficient large amplitude it will start affecting the Rossby
wave. Now in a reduced order model only resolving the Rossby wave but
not the synoptic wave this interaction cannot be explicitly
represented. However, because the Rossby wave initially triggered the
synoptic wave which then in turn affects it some days later, this can
be modeled with memory terms which takes into account that the Rossby
wave has triggered at time $t_0$ an anomaly which will affect it at some
later time $t_n$.

Recently, Wouters and Lucarini \cite{Wouters:2012} have proposed to
treat comprehensively the problem of model reduction in multi-scale
systems by adapting  the Ruelle response theory
\cite{Ruelle:1999,Ruelle:2009} for studying the effect of the coupling
between the fast and slow degrees of freedom of the system. This
theory has previously been used in a geophysical context to study the
linear and nonlinear response to perturbations
\cite{Lucarini:2009,Lucarini:2011}, which also allows climate change
predictions. This approach is based on the {\it chaotic hypothesis}
\cite{Gallavotti:1995} and allows the general derivation of the
reduced dynamics of the slow variables able to mimic the effect of the
fast variables in terms of matching the changes in the expectation
values of the observables of the slow variables. The ensuing
parametrization includes a deterministic correction, which is a mean
field result and corresponds to linear response, a general correlated
noise and a non-Markovian (memory) term. These results generalize Eq. (2). In the limit of
infinite time-scale separation, the classical results of the averaging
method is recovered. Quite reassuringly, the same parametrizations can
be found using a classical Mori-Zwanzig approach \cite{Chorin:2013}, which is
based on projecting the full dynamics on the slow variables and general
mathematical results provide evidence that deterministic, stochastic and
non-Markovian components should constitute the backbone of parameterizations 
\cite{Wouters:2013,Chekroun:2013}. Recent studies show improvements over
approaches based on time-scale separation
\cite{Wouters:2012,Wouters:2013,Chekroun:2013}.

Recent studies have shown that stochastic approaches are also
important for the prediction of extreme events and tipping points
\citep{Sura:2011,Sura:2013,Franzke:2012,Franzke:2013}. \citet{Sura:2013}
discusses a stochastic theory of extreme events. He especially focuses
on deviations from a Gaussian distribution; i.e. skewness and
kurtosis, as first measures of extremes. He shows that multiplicative
noise plays a significant role in causing non-Gaussian
distributions. \citet{Franzke:2012} shows that both deterministic
nonlinearity and multiplicative noise are important in predicting of
extreme events.

\section{Statistical Physics Approaches to Stochastic Climate
  Theory}
Significant progress has been achieved in the
development of tractable and accurate statistical dynamical closures
for general inhomogeneous turbulent flows that are underpinned by
exact stochastic models (see Box 2). For an accessible review see the text
book by \citet{Heinz:2003}. The statistical dynamical closure
theory, pioneered by \citet{Kraichnan_1959}, has been recognized
as a natural framework for a systematic approach to modelling
turbulent geophysical flows. Closure theories like the Direct
Interaction Approximation (DIA), \citep{Kraichnan_1959} for
homogeneous turbulence and the Quasi-Diagonal Direct Interaction
Approximation (QDIA), \citep{Frederiksen_1999} for the interaction of
mean flows with inhomogeneous turbulence have exact generalized
Langevin model representations \citep{Herring_Kraichnan_1972}. This
means that such closures are realizable; i.e. they have non-negative
energy.

The first major application of turbulence closures has been the
examination of the predictability of geophysical flows. Early
approaches applied homogeneous turbulence models to predicting error
growth \citep{Lorenz_1965,Leith_1971,LeithKraic_1972} whereas more
recent advances by \citet{FO_2005,OKane_Frederiksen_2008b}, building on the
pioneering studies of \citet{Epstein69} and
\citet{Pitcher77}, have enabled predictability studies of inhomogeneous strongly non-Gaussian flows typical of the mid-latitude atmosphere. Turbulence closures have also been used for
Subgrid-Scale Parameterisation (SSP) of the unresolved scales, for
example eddies in atmospheric and ocean general circulation models. 
Since it is generally only possible to represent the statistical
effects of unresolved eddies while their phase relationships with the
resolved scales are lost \citep{McComb_etal_2001}, statistical
dynamical turbulence closures are sufficient to allow SSPs to be formulated in a
completely transparent way
\citep{Leith_1971,Kraichnan_1976,Rose_1977,Leslie_Quarini_1979,Frederiksen_Davies_1997,Frederiksen_1999,OKane_Frederiksen_2008a}. Insights gained through the
development of inhomogeneous turbulence closure theory have motivated the recent development of general stochastic forms for subgrid-scale parameterisations for geophysical flows \citep{KFZ2012}.

\subsection{Statistical dynamical closure theory}
\noindent\fbox{
\parbox{\textwidth}{
{\bf BOX 2}\\
CLOSURE PROBLEM\\

In order to describe the statistical behavior of a turbulent flow the
underlying nonlinear dynamical equations must be averaged. For
simplicity we consider a generic equation of motion with quadratic
nonlinearity for homogeneous turbulence, in which the mean field is
zero, and the fluctuating part of the vorticity in Fourier space,
$\hat{\zeta_{\bf k}}$, satisfies the equation: 
\begin{eqnarray}
\frac{\partial}{\partial t}\hat{\zeta}_{\bf k}(t)=K_{\bf kpq}\hat{\zeta}_{-{\bf p}}(t)\hat{\zeta}_{-{\bf q}}(t).
\end{eqnarray}
where {$\bf p$} and ${\bf q}$ are the other wave numbers describing
triad interactions i.e. ${\bf k}=(k_{x},k_{y})$ where $\delta({\bf
k}+{\bf p}+{\bf q})=1$ if ${\bf k}+{\bf p}+{\bf q}=0$ and $0$
otherwise. Here $K_{\bf kpq}$ are the interaction or mode coupling
coefficients. The correlation between the eddies can now be
represented by an equation for the covariance (cumulant in terms of wavenumbers ${\bf k}$ and ${\bf l}$) which is found to depend
on the third order cumulant in Fourier space:
\begin{eqnarray}
\frac{\partial}{\partial t}\langle\hat{\zeta}_{\bf k}(t)\hat{\zeta}_{-{\bf l}}(t')\rangle=
K_{\bf kpq}\langle\hat{\zeta}_{-{\bf p}}(t)\hat{\zeta}_{-{\bf q}}(t)\hat{\zeta}_{-{\bf l}}(t')\rangle.
\label{idealcovariance}
\end{eqnarray}
Similarly the third order cumulant depends on the fourth order and so on such
that we see that an infinite hierarchy of moment or cumulant equations is
produced. Statistical turbulence theory is principally concerned with the
methods by which this moment hierarchy is closed and the subsequent dynamics
of the closure equations. The fact that for homogeneous turbulence the
covariance matrix is diagonal greatly simplifies the problem. The majority of
closure schemes are derived using perturbation expansions of the nonlinear
terms in the primitive dynamical equations. The most successful theories
use formal renormalization techniques \cite{Heinz:2003,Chorin:2013}. 
}
}

The development of renormalized turbulence closures was pioneered by
Kraichnan's Eulerian \textsc{dia} \citep{Kraichnan_1959} for homogeneous
turbulence. The \textsc{dia}, so named because it only takes into account
directly interacting modes, can be readily regularised to include
approximations to the indirect interactions
\cite{Frederiksen_Davies_2000,OKane_Frederiksen_2004} which are required to
obtain the correct inertial range scaling laws. Other homogeneous closures
such as Herring's self consistent field theory (\textsc{scft}
\citep{Herring_1965}) and McComb's local energy transfer theory
(\textsc{let}), \citep{McComb_1974} were independently developed soon
after. The \textsc{dia}, \textsc{scft} and \textsc{let} theories have since
been shown to form a class of homogeneous closures that differ only in whether
and how a fluctuation dissipation theorem (\textsc{fdt} i.e. the linear
response of a system to an infinitesimal perturbation as it relaxes toward
equilibrium)
\citep{Kraichnan_1959,Deker_Haake_1975,Carnevale_Frederiksen_1983,FDB_1994} is
applied. As noted earlier, a consequence of the \textsc{dia} having an exact
stochastic model representation is that it is physically realizable, ensuring
positive energy spectra. This is in contrast with closures based on the
quasi-normal hypothesis which require further modifications in order to ensure
realizability; an example of such a closure is the eddy damped quasi-normal
Markovian (EDQNM) closure \citep{Ogura_1963,Orszag_1970,Leith_1971} developed
as a bets Markovian fit to the DIA. The EDQNM is dependent on a choice of an
eddy-damping parameter which can be tuned to match the phenomenology of the
inertial range. This Markovian assumption assumes that the rate at which the
memory integral decays is significantly faster than the time scale on which
the covariances evolve. The relative success of these turbulence closures has
enabled the further study of the statistics of the predictability of
homogeneous turbulent flows
\citep{Leith_1971,Leith_1974,LeithKraic_1972,MetLes_1986}.

\citet{Frederiksen_1999} formulated a computationally tractable
non-Markovian (memory effects) closure, the quasi-diagonal direct interaction
approximation (\textsc{qdia}), for the interaction of general mean and
fluctuating flow components with inhomogeneous turbulence and
topography. The \textsc{qdia} assumes that, prior to renormalisation, a perturbative expansion of the covariances are diagonal at zeroth order. In general, very good agreement has been found between the
\textsc{qdia} closure results and the statistics of \textsc{dns}
\cite{Frederiksen_Davies_2000,OKane_Frederiksen_2004,FO_2005}. 

The non-Markovian closures discussed above are systems of
integro-differential equations with potentially long time-history
integrals posing considerable computational
challenges, however various ways to overcome these challenges exist
\cite{Rose_1985,FDB_1994,OKane_Frederiksen_2004,FO_2005,OKane_Frederiksen_2008a,OKane_Frederiksen_2008b,OKane_Frederiksen_2008c,OKane_Frederiksen_2010}
and have been generalised to allow computationally tractable closure
models for inhomogeneous turbulent flow over topography to be
developed \cite{Frederiksen_1999,OKane_Frederiksen_2004,FO_2005}. An
alternative derivation of a stochastic model of the Navier-Stokes equations
has been put forward by \citet{Memin:2014}. It is based on a decomposition of
the velocity fields into a differentiable drift and a stochastic component.

\subsection{Statistical dynamical and stochastic subgrid modelling}

Many subgrid-scale stress models assume the small scales to be close
to isotropic and in equilibrium such that energy production and
dissipation are in balance, similar to the eddy viscosity assumption
of the Smagorinsky model \citep{Smagorinsky_1963}. Using the
\textsc{dia}, \citet{Kraichnan_1959} showed that for
isotropic turbulence the inertial transfer of energy could be
represented as a combination of both an eddy viscous (on average
energy drain from retained to subgrid scales) and stochastic
back-scatter (positive semi-definite energy input from subgrid to
retained scales) terms. The nonlinear transfer terms represented by
eddy viscosity and stochastic back-scatter are the subgrid processes
associated with the respective eddy-damping and  nonlinear noise terms
that constitute the right hand side of the \textsc{dia} tendency
equation for the two-point cumulant $\frac{\partial}{\partial t}\langle\hat{\zeta}_{\bf k}(t)\hat{\zeta}_{-{\bf k}}(t')\rangle$. \citet{Leith_1971} used the \textsc{edqnm} closure to
calculate an eddy dissipation function that would preserve a
stationary $k^{-3}$  kinetic energy spectrum for two-dimensional
turbulence. \citet{Kraichnan_1976} developed
the theory of eddy viscosity in two and three dimensions and was the
first to identify the existence of a strong cusp in the spectral eddy
viscosity near the cutoff wavenumber representing local interactions
between modes below and near the cusp. \citet{Rose_1977} argued for
the importance of eddy noise in subgrid modelling. 

\citet{OKane_Frederiksen_2008a} calculated
\textsc{qdia} based \textsc{ssp}s considering observed atmospheric
flows over global topography and quantifying the relative importance
of the subgrid-scale eddy-topographic, eddy-mean field, quadratic mean
and mean field-topographic terms. They also compared the \textsc{qdia}
based \textsc{ssp}s to heuristic
approaches based on maximum entropy, used to improve systematic
deficiencies in ocean climate models \citep{Holloway92}. 
While closure models may be the natural starting place for developing
subgrid-scale parameterisations, their complexity makes them difficult
to formulate and apply to multi-field models like
\textsc{gcm}s, even though sucessfull studies exist
\cite{ZidFred2009,KFZ2012}. 

\section{Stochastic Parametrisation Schemes in Comprehensive Models}
Climate and weather predictions are only feasible because the governing
equations of motion and thermodynamics are known. To
solve these equations we need to resort to numerical simulations that
discretize the continuous equations onto a finite grid and parameterize all
processes that cannot be explicitly resolved. Such models can be
characterized in terms of their dynamical core, which describe the resolved
scales of motion, and the physical parameterizations, which provide estimates
of the grid-scale effect of processes which cannot be resolved by the
dynamical core. This general approach has been hugely successful in that
nowadays predictions of weather and climate are made routinely. On the other
hand, exactly through these predictions it has become apparent that
uncertainty estimates produced by current state-of the art models
still have shortcomings.

One shortcoming is that many physical parameterizations are based on bulk
formula which are based on the assumption that the subgrid-scale state
is in equilibrium with the resolved state \cite{Pa01}. Model errors
might arise from a misrepresentation of unresolved subgrid-scale
processes which can affect not only the variability, but also the mean
error of a model
\cite{Sardeshmukh:2001, Penland:2003}. 
An example in a comprehensive climate model is e.g., 
the bias in the 500hPa geopotential height pattern,
which is reduced when the representation of the subgrid-state is refined \cite{Be12}
(Fig.\,2).

In recent years, methods for the estimation of flow-dependent uncertainty in
predictions have become an important topic. Ideally, uncertainties
should be estimated within the physical parameterizations and
uncertainty representations should be developed alongside the
model. Many of these methods are ``ad hoc'' and added {\it a
  posteriori} to an already tuned model. Only first steps to develop
uncertainty estimates from within the parameterizations have been
attempted \cite{Cohen:2006,Plant:2008}.

The representation of model-error in weather and climate models falls in one
of two major categories: Multi-model approaches and stochastic
parameterizations. In the multi-model approach each ensemble member consists
of an altogether different model. The models can differ in the dynamical core
and the physical parameterizations \cite{Kr00,Ha05,Ho96} or use the same
dynamical core but utilize either different static parameters in their
physical parameterizations \cite{St05} or altogether different physics
packages \cite{Murphy:2004,Stensrud:2000,Eckel:2005,Be11}.
Both approaches have been successful in improving predictions of weather 
and climate over a range of scales, as well as their uncertainty.
Multi-model ensembles provide more reliable seasonal forecasts \cite{palmer2004development}
and are commonly used for the uncertainty assessment of climate change predictions 
e.g., as in the Assessment Report 5 of the Intergovernmental Panel on Climate
Change (IPCC) 
\cite{IPCC13}.
Stochastic parameterizations are routinely used to improve the reliability of weather 
forecasts in the short- \cite{Be11} and medium-range 
\cite{Bo09,Be09,Pa09} as well as for seasonal predictions \cite{Be08,Do09,weisheimer2011assessment}.

In the stochastic approach, the effect of uncertainties due to the finite
truncation of the model are treated as independent realizations of stochastic
processes that describe these truncation uncertainties. This treatment goes
back to the idea of stochastic-dynamic prediction
\cite{EpsteinPitcher1972,Pitcher77,Pa01}. While the verdict is still
open if subgrid-scale fluctuations must be included explicitly via a
stochastic term, or if it is sufficient to include their mean
influence by improved deterministic physics parameterizations, one
advantage of stochastically perturbed models is that all ensemble
members have the same climatology and model bias; while for
multi-parameter,  multi-parameterization and multi-model ensembles
each ensemble member is {\it de facto} a different model with its own
dynamical attractor. For operational centers the maintenance of
different parameterizations requires additional resources and due to the different biases
makes post-processing very difficult.

\subsection{Stochastic Parameterizations in Numerical Weather
  Prediction}
Due to the chaotic nature of the dynamical equations governing the
evolution of weather, forecasts are sensitive to the initial condition
limiting the intrinsic predictability of the weather system
\cite{Lorenz:1963,Lorenz:1969}. Probabilistic forecasts are performed
by running ensemble systems,  where each member is initialized from a
different initial state and much effort has gone into the optimal
initialization of such ensemble systems
\cite{MoPa93,ToKa93,Ho96}. Nevertheless state-of-the-art numerical
weather predictions systems continue to produce  unreliable and
over-confident forecasts \cite{Buizza:2005}. Consequently, the other
source of forecast uncertainty -- model-error -- has received
increasing attention \cite{Palmer:1999,Pa01}. Since for chaotic systems
model-error and initial condition error will both result in
trajectories that will diverge from the truth, it is very difficult to
disentangle them \cite{smith2001disentangling}.

The first stochastic parameterization used in an operational numerical
weather prediction model was the stochastically perturbed physics
tendency scheme (SPPT), sometimes also referred to as stochastic
diabatic tendency or  Buizza-Miller-Palmer (BMP) scheme
\cite{Buizza:1999}. SPPT is based on the notion that -- especially as
the horizontal resolution increases --  the equilibrium assumption no
longer holds and the subgrid-scale state should be sampled rather than
represented by the equilibrium mean. Consequently, SPPT multiplies the
accumulated physical tendencies  of temperature, wind and moisture at
each grid-point and time step with a random pattern that has spatial
and temporal correlations. In other words, SPPT assumes that
parameterization uncertainty  can be expressed as a multiplicative
noise term. Ensemble systems perturbed with the SPPT scheme show
increased probabilistic skill  mostly due to increased spread in short
and medium-range ensemble forecasts
\cite{Buizza:1999,Teixera:2008,Reynolds:2011,Be14}.

A second successful stochastic parameterization scheme, is the
so-called stochastic kinetic energy-backscatter scheme (SKEBS) whose origin
lies in Large-Eddy Simulation modeling \cite{MaTh92} and has recently
been extended to weather and climate scales \cite{Sh05,ShPa03}. The
key idea is that energy associated with subgrid processes is injected back onto the grid
using a stochastic pattern generator. This method has been
successfully used in a number of operational and research forecasts
across a range of scales
\cite{Be09,Be11,Be14,Bo08,Bo09,Ch10,Sa14}. Similar to the SPPT scheme,
ensemble systems with SKEBS increase probabilistic skill by increasing
spread and decreasing the root-mean-square (RMS) error of the ensemble
mean forecast. First results of these schemes at a convection-permitting
resolution of around 4km report also a positive impact on forecast skill, in
particular more reliable precipitation forecasts \cite{Bou12,Ro14}.

\subsection{Stochastic Parameterizations in Climate Models}
The use of stochastic parameterization in climate models is still in
its infancy. Climate prediction uncertainty assessments, e.g., IPCC \cite{IPCC:2013}, are  almost exclusively based on
multi-models, mostly from different research centers.  Part of the
problem is that on climate timescales, limited data for verification
exists. A second reason is that on longer timescales, bias is a major
source of uncertainty and traditional multi-models are very efficient
at sampling biases, although such an experiment is poorly designed for
an objective and reliable uncertainty assessment.

\begin{figure}
\begin{center}
%\mbox{\makebox[0pt]{\large a)}\parbox[t][][b]{0.5\textwidth}{
%\psfig{fig2.pdf,bb=0 0 575 540,clip=,angle=0,width=0.46\textwidth}
%}
\noindent\includegraphics[width=1.0\textwidth]{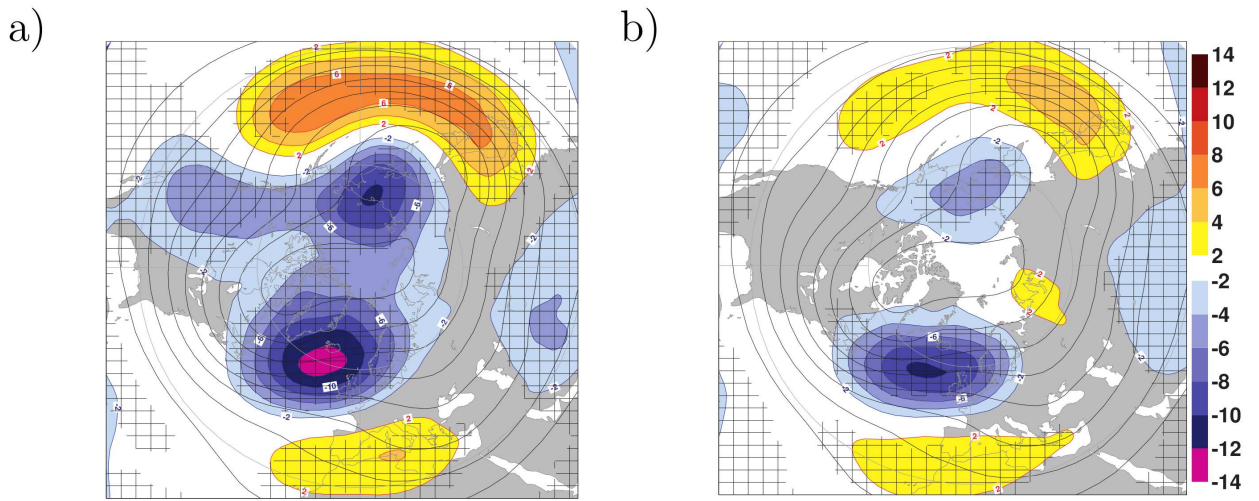}
\end{center} 
\caption{\label{fig2}
Mean systematic error of 500 hPa geopotential height fields (shading) for
extended boreal winters (December--March) of the period 1962-2005. Errors are
defined with regard to the observed mean field (contours), consisting of a
combination of ERA-40 (1962-2001) and operational ECMWF analyses
(2002-2005). (a) Systematic error in a numerical simulation with the ECMWF
model IFS, version CY32R1, run at a horizontal resolution of $T_L95$ (about
210km) and 91 vertical levels. (b) Systematic error in a simulation with a
stochastic kinetic-energy backscatter scheme (SKEBS). Significant differences
at the 95\% confidence level based on a Student's t-test are hatched. After
Berner et al. \cite{Be12}.
\label{fig_z500}
}
\end{figure}

However, in recent years first studies have emerged which demonstrate
the ability of stochastic parameterizations to reduce longstanding
biases and improve climate variability (see Fig. \ref{fig2} for an
example). Jin and Neelin \cite{LiNe00} developed a stochastic
convective parameterization that includes a random contribution to the
convective available potential energy(CAPE) in the deep convective
scheme. They find that adding convective noise results in enhanced
eastward propagating, low-wavenumber low-frequency
variability. \citet{Be12} investigate the impact of SKEBS on
systematic model-error and report an improvement in the representation
of convectively-coupled waves leading to a reduction in the tropical
precipitation bias. Furthermore, Majda and colleagues developed
systematic stochastic multi-cloud parameterizations for organized
convection
\cite{Majda:2008,Majda:2008b,Khouider:2010,Frenkel:2011}. The
multi-cloud approach is based on the assumption that organized
convection involves three types of clouds and the evolution from one
cloud type to another can be described by a transition matrix.

A longstanding systematic error of climate models is the underestimation of
the occurrence of Northern Hemispheric blocking. Stochastic
parameterizations have been demonstrated to be one way to increase
their frequency  \cite{Ju05b,Be08,Do09,weisheimer2011assessment},
although, e.g. increasing horizontal resolution, leads to similar
improvements \cite{Be12, Dawson:2012}. This suggests that while it
might be necessary to include subgrid-scale variability in some form,
the details of this representation might not matter. On the other
side, \citet{Be12} argue that this degeneracy of response to different
subgrid-scale forcings warrants a cautionary note: namely that a
decrease in systematic error might not necessarily occur for the right
dynamical reasons. 
The opposite holds true, as well: Due to the necessary tuning of parameters 
in the parameterizations of comprehensive climate 
models, an improvement in the formulation of a physical process might not 
immediately lead to an improved model performance.
A striking example of compensating model errors is described in 
\citet{PaWe11}, who report
how an inadequate representation of horizontal baroclinic fluxes resulted in a model error that 
was equal and opposite to the systematic error caused by insufficiently 
represented vertical orographic gravity wave fluxes. Improvements to wave 
drag parameterization without increasing resolution unbalanced the 
compensating model errors, leading to an increase in systematic model bias. 

Williams \cite{williams2012climatic} studied the effect of including a
stochastic term in
the fluxes between the atmospheric and oceanic components in a coupled
ocean-atmosphere model. He reports  changes to the time-mean climate
and increased variability of the El Nino Southern Oscillation,
suggesting that the lack of representing of sub-grid variability in
air-sea fluxes may contribute to some of the biases exhibited by
contemporary climate models. 

On seasonal timescales where sufficient observational data for a
probabilistic verification exist, stochastic parameterizations have
been reported to increase predictive skill. For example, ensemble
forecasts of the sea surface temperatures over the  Nino3.4 region
showed increased anomaly correlation, decreased bias and decreased
root mean square error in coupled ocean-atmosphere models
\cite{Be08,Do09,weisheimer2011assessment}.

{\noindent\section{Conclusion}
\label{sec:conclusion}}

% Postulate the use of Stochastic strategies
We postulate the use of stochastic-dynamical models for uncertainty assessment
and model-error representation in comprehensive Earth-System models. This need
arises since even state-of-the-art weather and climate models cannot resolve
all necessary processes and scales. Here we reviewed mathematical
methods for stochastic climate modeling as well as
stochastic subgrid-scale parameterizations and postulate their use for
a more systematic strategy of parameterizing unresolved and
unrepresented processes.

% Evidence that ad hoc stochastic parameterizations work
In the last decade, a number of studies emerged that demonstrate the potential
of this approach, albeit applied in an ad hoc manner and
tuned to specific applications. Stochastic parameterizations have been
shown to provide more skillful weather forecasts than traditional
ensemble prediction methods, at least on timescales where verification
data exists. In addition, they have been shown to reduce longstanding
climate biases, which play an important role especially for climate
and climate change predictions.

% Predicting uncertainty
Here we argue, that rather than pushing out the limit of skillful ensemble
predictions by a few days, more attention should be given on the assessment of
uncertainty (as already proposed, e.g.,
\citet{smith2001disentangling}). Ideally, it should be carried out alongside
the physical parameterization and dynamical core development and not added
{\it a posteriori}. The uncertainty should be directly estimated from within
the parameterization schemes and not tuned to yield a particular model
performance, as is current practice. For example, \citet{Sapsis:2013a,Sapsis:2013b}
propose a statistical framework which systematically quantifies
uncertainties in a stochastic fashion.

% Limit of Predictions/Predictability
The fact that according to the last two assessment reports (AR) of the
IPCC (AR4 \cite{IPCC:2007} and AR5 \cite{IPCC:2013}) the
uncertainty in climate predictions and projections has not decreased may be
a sign that we might be reaching the limit of climate predictability,
which is the result of the intrinsically nonlinear character of the climate
system (as first suggested by \citet{Lorenz:1963}).

Recently Palmer \cite{palmer2014more} argued that due to limited computational
and energy power resources, predictable scales should be solved accurately,
while the unpredictable scales could be represented inaccurately. This
strategy is at the core of the systematic mode reduction reviewed here, but
has only recently been considered for comprehensive Earth-System
Models. Stochastic models focus on the accurate simulation of the large,
predictable, scales, while only the statistical properties of the small,
unpredictable, scales are captured. This has been demonstrated, e.g,
by \citet{Franzke:2006,Kravtsov:2005}, who successfully applied mode
reduction strategies to global atmospheric circulation models. They
showed that these reduced models consisting of only 10-15 degrees of
freedom reproduced many of the important statistics of the numerical
circulation models which contained a few hundreds degrees of
freedom. \citet{Vandeneijnden:2003} proposed numerical approaches for
multi-scale systems where only the largest scales are explicitly
computed and the smaller scales are approximated on the fly.

The recent result of Wouters and Lucarini
\cite{Wouters:2012,Wouters:2013} provide a promising path towards a
general theory of parametrizations for weather and climate models, and
give theoretical support that parameterization schemes should include
deterministic, stochastic and non-Markovian (memory)
components. Moreover, Wouters and Lucarini's results suggest that
there is common ground in developing parameterizations for weather and
climate prediction models. Optimal representations of the reduced
dynamics based on Ruelle's response theory and the Mori-Zwanzig
formalism coincide, thus, providing equal optimal representations of
the long-term statistical properties and the finite-time evolution of
the slow variables.

One exciting future research area is the use of stochastic methods for use in
data assimilation, which is already an active field of research
\citep{Milleretal99,isaksen2007use,OKane_Frederiksen_2008c,OKane_Frederiksen_2010,
Ha14,Ro14,Gottwald:2013a}. Stochastic methods have been shown to increase the ensemble
spread in data assimilation, leading to a better match between observations
and model forecasts \cite{Mitchell:2012, Ha14, Ro14}. A cutting-edge frontier is the
use of order moments and memory effects in Kalman filter data
assimilation schemes \citep{OKane_Frederiksen_2010}. Another emerging field is
the use of stochastic parameterizations in large climate ensembles, which
would allow the comparison of uncertainty estimates based in multi-models to
that of stochastically perturbed ones.

Our hope is that basing stochastic-dynamic prediction on sound
mathematical and statistical physics concepts will lead to substantial
improvements, not only in our ability to accurately simulate weather
and climate, but even more importantly to give proper estimates on the
uncertainty of these predictions.

\noindent{\bf Acknowledgments:}\\
CLEF is supported by the German Research Foundation (DFG) through the
cluster of Excellence CliSAP, TJO is an
Australian Research Council Future Fellow, PDW acknowledges a
University Research Fellowship from the Royal Society (UF080256) and
VL funding from the European Research Council (NAMASTE).

\bibliographystyle{unsrtnat}

\clearpage

\listoffigures

\clearpage

\clearpage

\end{document}